\documentclass[article]{aa}
\usepackage{natbib}
\usepackage{multirow}
\bibpunct{(}{)}{;}{a}{}{,}
 \usepackage[dvips]{graphicx}

\DeclareGraphicsExtensions{.ps}                      
\usepackage{txfonts}
\begin{document}
%
    \title{Performance of the VLT Planet Finder SPHERE II.} 
     \subtitle{Data analysis and Results for IFS in laboratory}


   \author{D. Mesa\inst{1}, R. Gratton\inst{1}, A. Zurlo\inst{1,2}, 
    A. Vigan\inst{2}, R.U. Claudi\inst{1}, \\
    M. Alberi\inst{3},  J. Antichi\inst{4}, A. Baruffolo\inst{1}, 
    J.-L. Beuzit\inst{5}, A. Boccaletti\inst{6}, M. Bonnefoy\inst{5}, A. Costille\inst{2,5}, S. Desidera\inst{1}, 
    K. Dohlen\inst{2}, D. Fantinel\inst{1}, M. Feldt\inst{7}, T. Fusco\inst{2}, E. Giro\inst{1},
    T. Henning\inst{7}, M. Kasper\inst{8}, M. Langlois\inst{9}, A.-L. Maire\inst{1}, P. Martinez\inst{10}, 
    O. Moeller-Nilsson\inst{7}, D. Mouillet\inst{5}, C. Moutou\inst{2}, A. Pavlov\inst{7}, 
    P. Puget\inst{5}, B. Salasnich\inst{1}, J.-F. Sauvage\inst{5,11}, E. Sissa\inst{1}, M.Turatto\inst{1}, 
    S. Udry\inst{12}, F. Vakili\inst{10}, R. Waters\inst{13}, F. Wildi\inst{12} 
          }

   \institute{\inst{1}INAF-Osservatorio Astronomico di Padova, Vicolo dell'Osservatorio 5, Padova, Italy, 35122-I \\
   \inst{2}Aix Marseille Universit\'e, CNRS, LAM - Laboratoire d'Astrophysique de Marseille, UMR 7326, 13388, Marseille, France\\
   \inst{3} Dipartimento di Fisica ed Astronomia - Universit\'a di Bologna, Viale Berti Pichat 6/2, I-40127 Bologna, Italy\\
   \inst{4}INAF - Osservatorio Astrofisico di Arcetri - L.go E. Fermi 5, I-50125 Firenze, Italy\\
   \inst{5}UJF-Grenoble 1 / CNRS-INSU, Institut de Plan\'etologie et d'Astrophysique de Grenoble (IPAG) UMR 5274, Grenoble, F-38041, France\\ 
    \inst{6}LESIA, Observatoire de Paris, CNRS, University Pierre et Marie Curie Paris 6 and University Denis Diderot 
         Paris 7, 5 place Jules Janssen, F-92195 Meudon, France\\
    \inst{7}Max-Planck-Institut f\"ur Astronomie, K\"onigstuhl 17, 69117 Heidelberg, Germany\\
    \inst{8}European Southern Observatory, Karl-Schwarzschild-Strasse 2, D-85748 Garching, Germany\\
    \inst{9}CRAL, UMR 5574, CNRS, Universit\'e Lyon 1, 9 avenue Charles Andr\'e, 69561 Saint Genis Laval Cedex, France\\
    \inst{10}Laboratoire J.-L. Lagrange, UMR 7293, Observatoire de la C\^ote d’Azur (OCA), Universit\'e de Nice-Sophia Antipolis (UNS),
         CNRS, Campus Valrose, 06108 Nice Cedex 2, France\\
    \inst{11}ONERA - The French Aerospace Lab BP72 - 29 avenue de la Division Leclerc FR-92322 Chatillon Cedex\\
    \inst{12}Observatoire de Gen\`eve, University of Geneva, 51 Chemin des Maillettes, 1290, Versoix, Switzerland\\
    \inst{13}Sterrenkundig Instituut Anton Pannekoek, University of Amsterdam, Science Park 904, 1098 Amsterdam, The Netherlands
    }

   \date{Received  / accepted }


  \abstract
   {}
   {We present the performance of the Integral Field Spectrograph (IFS) of SPHERE, the high-contrast 
   imager for the ESO VLT telescope designed to perform imaging and 
   spectroscopy of extrasolar planets, obtained from tests performed at 
   the Institute de Plan\'etologie et d'Astrophysique de Grenoble facility during the integration phase of the instrument.}
   {The tests were performed using the instrument software purposely prepared for 
   SPHERE. The output data were reduced applying the SPHERE data reduction and handling 
   software, adding an improved spectral deconvolution procedure. To this aim, we prepared an alternative 
   procedure for the spectral subtraction exploiting the principal components analysis 
   algorithm. Moreover, a simulated angular differential imaging procedure was also 
   implemented to estimate how the 
   instrument performed once this procedure was applied at telescope. The capability of the 
   IFS to faithfully retrieve the spectra of the detected faint companions was also considered.}
   {We found that the application of the updated version of the spectral deconvolution procedure 
   alone, when the algorithm throughput is properly taken into account, gives us a 
   $5\sigma$ limiting contrast of the order of 5$\times$$10^{-6}$ or slightly better. The 
   further application of the angular differential imaging procedure on these data should allow us to improve the contrast by 
   one order of magnitude down to around 7$\times$$10^{-7}$ at a separation of 0.3 arcsec. The application of a 
   principal components analysis procedure that simultaneously uses spectral and angular data gives comparable results. 
   Finally, we found that the reproducibility of the spectra of the detected faint
   companions is greatly improved when angular differential imaging is applied in addition to the spectral deconvolution.}
   {}


\titlerunning{IFS performance}
\authorrunning{Mesa et al.}
   \maketitle
%

\section{Introduction}
High-contrast direct imaging of extrasolar planets is very challenging because of the very
high brightness contrast between the planets and their host ($10^{-6}$ for young giant planets with 
intrinsic emission observed in the near infrared  
and $10^{-8}-10^{-9}$ for old giant and rocky planets seen through stellar reflected light), and 
the very small angular separation between the planets and their hosts (few tenths of arcsec for planet at $\sim$10 AU from 
their hosts and tens of parsec away). For this reason, 
only 47 low-mass companions out of more than 1000 planets discovered so far 
have been detected by exploiting high-contrast imaging \citep[see Extrasolar Enciclopaedia\footnote{http://exoplanet.eu} - ][]
{Sch11}. A 
new generation of instruments devoted to exploiting this technique is now becoming operational. They
combine extreme adaptive optics (XAO), providing high Strehl ratio (SR) 
by correcting high-order wavefront aberrations, with high-efficiency coronagraphs to attenuate the 
diffracted light from the host star. 
To fully exploit the potential of these instruments, differential imaging techniques, such as
angular differential imaging (ADI; see \citealt{Ma06}), simultaneous spectral differential
imaging (S-SDI; see, e.g., \citealt{Rac99} and \citealt{Ma05}) and spectral deconvolution\footnote{
The authors are aware that the definition of spectral deconvolution is probably inadequate since
no deconvolution is actually performed. However, this name is routinely used in the high-contrast imaging community. Moreover,
the method is clearly explained in Section~\ref{specdeconvIFS}. For these reasons, we have decided to keep it in our paper.} (SD; see 
\citealt{Sparks02} and \citealt{Th07}) are also used. One of these instruments, Project 1640 
\citep{Crepp11}, is already operating at the 5 m Palomar Telescope and led to good scientific 
results (see, e.g., \citealt{Opp13}). 
The Gemini Planet Imager (GPI; see \citealt{Mac06}) is starting science operations at the Gemini South 
Telescope while SPHERE is currently in its science verification phase at Paranal. 
A fourth instrument called CHARIS is expected to become operative by 
the end of 2015~\citep{PetLim13} at the Subaru Telescope. This instrument is expected to expand
the capabilities of SCExAO, already available at the same telescope \citep[see e.g. -][]{Cur14}.
Besides these instruments, the LBT telescope with its lower order adaptive optics system has 
proven to be a very powerful instrument for the imaging of extrasolar planets \citep[see e.g.][]{Ske12,Esp13}.\par
The instrument SPHERE~\citep{Beu06}, dedicated to high-contrast imaging of faint planetary mass
companions around nearby young stars, is installed on the Nasmyth platform of the ESO Very Large Telescope (VLT). 
It is divided into four subsystems:
\begin{itemize}
\item the Common Path and Infrastructure (CPI) receives and transfers telescope light, 
feeds the other three subsystems with a highly stabilized adaptive optics (AO) corrected and
coronagraphic beam. The Extreme AO system for SPHERE, called SAXO, produces a beam 
with a Strehl ratio better than 90\%~\citep{Petit2012}. The AO deformable mirror (DM) is described in~\citet{Hugot2012}.
\item the Zurich IMaging POLarimeter (ZIMPOL) produces direct and differential polarimetric images in the visible
between 500 and 900 nm~\citep{Th08}.
\item the InfraRed Dual-band Imager and Spectrograph (IRDIS) has various observing modes: 
classical imaging (CI), dual-band imaging (DBI; see~\citealt{Vig10}), 
polarimetric imaging (DPI; see~\citealt{Lan10}), or long-slit spectroscopy (LSS - see~\citealt{Vig08}).
It works in the near infrared 
between 0.95 and 2.32 $\mu$m with a field of view (FOV) of 12.5$\times$11 arcsec~\citep{Do08}. 
\item the IFS also operates in the near infrared between 
0.95 and 1.35 $\mu$m with a 2-pixel spectral resolution of 50 for the YJ mode and between 0.95 and 1.65 $\mu$m 
with a spectral resolution of around 30 for the YH 
mode~\citep{Cl08}. Its FOV is of 1.7 arcsec square. The integral 
field unit (IFU) is composed of a double system of lenslets set on an a-focal 
configuration and tailored to reduce cross-talk between both adjacent lenslets and adjacent 
spectra \citep{Ant09}. The raw data for IFS is composed of 
around 21000 spectra aligned with the detector columns over a hexagonal grid rotated by 
10.9$^{\circ}$ with respect to the dispersion axis. The data reduction and handling (DRH) 
software~\citep{Pa08} transforms each detector frame into a datacube composed of 39 monochromatic 
images of 291x291 pixels for both the observing modes. 
\end{itemize}
Among the other modes of SPHERE, the IFS achieves very high contrast
by combining spectral and angular differential imaging.
In recent years several teams have dedicated efforts toward developing spectral extracting
tools for high-contrast imaging with integral field spectrographs data. 
For example, \citet{Sei07} and \citet{Lav09} presented an analysis aimed at extracting the spectrum of GQ Lup b. 
Moreover, \citet{Bow10} and \citet{Bar11} performed high-contrast spectroscopy of HR8799b. More recently, the spectroscopy 
of all HR8799 system planets was performed by \citet{Opp13} and a library of spectra
of 15 young M6-L0 dwarfs was compiled by \citet{Bon14}. A method for the
spectral characterization of faint companions exploiting the IFS of Projects 1640 was 
described in \citet{Pueyo2014} while the TLOCI algorithm \citep{Mar14} was developed to exploit
GPI IFS data, using both the ADI and the S-SDI. A method for high-contrast
imaging with ADI technique was developed in the context of the
SEEDS pipeline \citep{Bra13}. \par
In this paper we present the results of the laboratory tests performed and highlight the
performance limits of the IFS in terms of contrast and signal to noise (S/N) as determined from laboratory data. We 
obtained these results with dedicated speckle subtraction algorithms, which we will describe.
While on-sky data from SPHERE are currently becoming available we think that this paper exploiting the laboratory data is still
useful given that it allows us to present the state-of-the-art data reduction algorithms for SPHERE IFS.     
  
The outline of this paper is: in Section~\ref{IFStest} we describe the test 
procedures, in Section~\ref{datared} we describe the data reduction pipeline, in 
Section~\ref{result} and Section~\ref{resultyh} we present the results of the data reduction
for the YJ- and YH-mode, and in Section~\ref{conclusion} we give our 
conclusions.


\section{The IFS tests description}\label{IFStest}
The whole SPHERE instrument is at present in its science verification phase at Paranal. Before shipment
to ESO, it was assembled in a large dedicated clean room at Institute de Plan\'etologie et 
d'Astrophysique de Grenoble (IPAG). A large number of 
functional and performance tests were performed on the instrument from the end of 2011 to the end of 2013. 
The aim of this paper is to describe the tests performed on IFS. The first 
goal of these tests was to check that the instrument control software (INS) worked properly. 
It is organized in different components, defined as "templates", which can be run autonomously or within observing blocks, 
according to the operational model adopted at VLT \citep[see e.g. ][]{Bar12}. These templates execute all the tasks 
needed to perform the calibration of the instrument and acquire scientific observations. \par
The main calibration templates perform the following tasks: detector
dark and flat field acquisition, determination of  the spectra positions, wavelength calibration 
and finally acquisition of the IFU flat field. The IFU flat field
is required to calibrate different outputs from different lenslets composing the lenslet array.
Other calibration templates were executed to monitor the instrument
such as the gain and readout noise (RON), ghosts, distortion and  background. 
All these templates were intensively and repeatedly tested and, when 
necessary, modifications were made to improve their outcome or to correct errors 
during their execution. Other calibration templates dedicated to the observations (e.g., sky 
flat) will be fully tested when the instrument is installed at the telescope.\par
Two templates enable the acquisition of scientific data: one of them is for observations of 
single objects with IFS in the YJ mode IRDIFS, while the second is for the observations in the YH mode 
(IRDIFS\_EXT). 
All these templates run IFS in parallel with IRDIS. Several dichroic beamsplitters in the CPI, located after the coronagraphic focal mask 
and before the Lyot stops, allow for the feeding of both instruments simultaneously. IRDIS is used both 
for acquisition purposes, like fine coronagraphic centering and the noncommon path aberrations 
calibration, as well as for scientific purposes.\par
\begin{figure}
\begin{center}
\includegraphics[width=8.0cm]{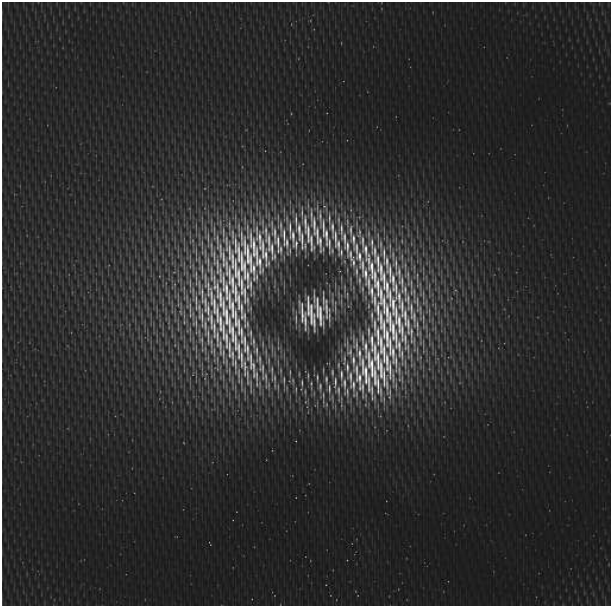}
\caption{Example of IFS coronagraphic raw data. The displayed image covers
a square 1.77$\times$1.77 arcsec FOV. This observation was performed in YJ-mode.\label{rawdata}}
\end{center}
\end{figure}
\begin{figure}
\begin{center}
\includegraphics[width=8.0cm]{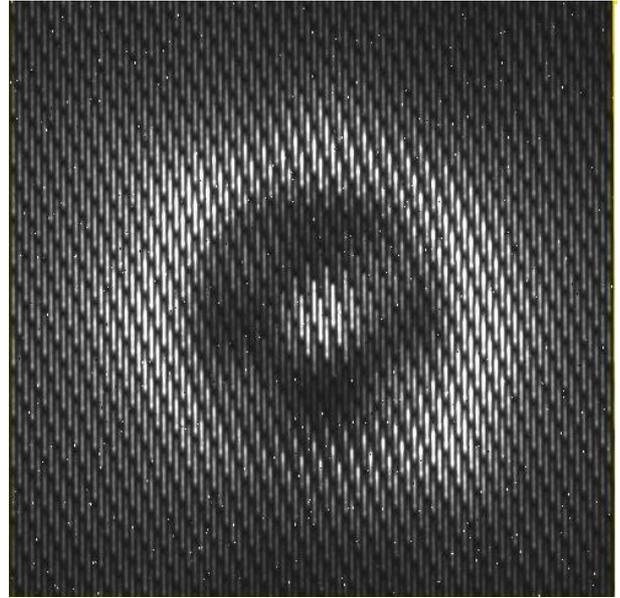}
\caption{Zoom on the central part of the image from Figure~\ref{rawdata} showing the spectra more clearly.\label{rawdatazoom}}
\end{center}
\end{figure}

The laboratory tests performed at IPAG used a telescope simulator (TSIM) including rotating phase screens 
to emulate atmospheric turbulence corresponding to 0.6\arcsec, 0.8\arcsec, and 1.2\arcsec seeing conditions. The 
phase screens have adjustable rotation speed to simulate wind speeds ranging from 2 to 10 m/s. 
Stars of different brightness were emulated by inserting different neutral density filters. 
It is important to say that this method just simulates a fainter star from the IFS point of view, while this is 
not true for the AO. This is important given that the AO performance degrades for faint stars.
Moreover, several coronagraphs (SPHERE allows the use of both apodized Lyot and 4-quadrant 
coronagraphs; see~\citealt{Boc08}) have been tested. However, we report on results obtained only with the apodized Lyot 
coronagraph with a mask diameter of 185 mas (projected on sky) since the performance of the 4-quadrant coronagraph is not yet optimized. 
Possible explanations 
for performances loss, compared to expectations, with this type of coronagraph could be
related to the presence of low-order aberrations, probably a defocus, or a possible aging of the optical components materials. 
These reasons could explain the higher level of performances obtained during the preliminary tests 
performed at LESIA \citep[see e.g.][]{Boc08}.
Figure~\ref{rawdata} shows an example of IFS (in the YJ-mode) coronagraphic raw data, while
Figure~\ref{rawdatazoom} shows a zoom on the central part of the same image highlighting spectral layout. 
In this case, we used a simulated star with a J magnitude of 2.6 with typical seeing 
and wind speed condition for Paranal (0.8 arcsec seeing and 5 m/s wind speed).

\section{Data reduction pipeline}\label{datared}
We performed the data reduction using the DRH 
pipeline developed by the SPHERE consortium. This software is composed by different modules, called {\em recipes},  
aimed to reduce specific data sets generated by the SPHERE
instrument software. The data reduction pipeline includes the following steps:
\begin{itemize}
\item creation of the master dark;
\item creation of the detector master flat field;
\item definition of the IFS spectra positions. At this step, each detector pixel is attached
to a particular spectrum and a first guess wavelength is assigned to it, based on a model. 
The corresponding data set is obtained by illuminating the IFS with a broadband light source 
connected to an integrating sphere located in the calibration arm;
\item wavelength calibration of each spectra. At this stage the wavelength assigned to each 
pixel is refined using data obtained illuminating the IFS with several laser light sources 
(three for the YJ-mode and four for the YH-mode). Those lasers feed the same integrating sphere.
\item creation of IFU flat field to take the response to a uniform illumination of all the IFU 
lenslets properly into account;
\item reduction of the scientific images with the creation of monochromatic images datacubes. 
The datacubes are obtained by interpolating the individual spectra spectrally and spatially.
A more detailed description of the procedure used in the DRH to create the calibrated datacube can be found in~\citet{Pa08};
\item subtraction of the speckle pattern by exploiting the SD method. 
\end{itemize}

To improve the operation of the instrument and the accuracy of the calibrations,
we requested upgrades of the DRH recipes in recent months.
Starting from the basis offered by the DRH pipeline, we also improved the spectral deconvolution 
procedure with the aim to replicate the results of the IFS performance simulations \citep{Mesa11}
and to improve the overall instrument performances. In particular, we improved the
algorithms for the definition of the star center, the wavelength rescaling 
factor, and of the full width at the half maximum (FWHM) of the speckles at different wavelengths. 
In addition, the principal component analysis (PCA) technique has been
adapted for the IFS. Both these procedures were developed using the IDL 
language. Their inclusion in the DRH pipeline is foreseen in the next years. In the following we describe these procedures in detail.
We would like to emphasize that the procedures described here are the results of a
series of different attempts aimed to optimize their performance. In the following we 
will describe the current versions of the procedures.

\subsection{Spectral deconvolution for IFS}\label{specdeconvIFS}
\begin{figure}
\begin{center}
\includegraphics[width=8.0cm]{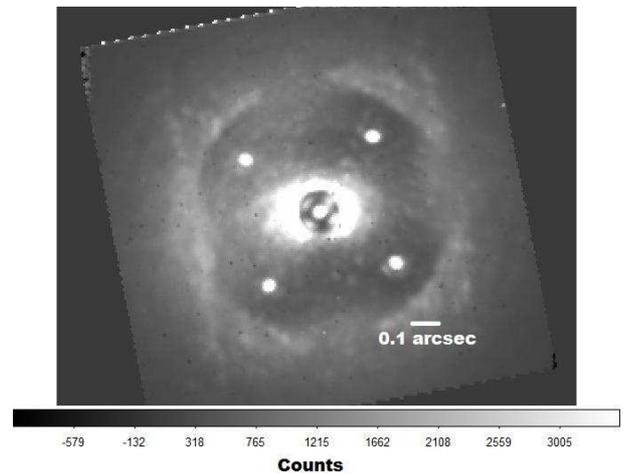}
\caption{Example of a monochromatic ($\lambda = 1.047 \mu m$) reduced image where the four 
artificial satellite spots are clearly visible. This image was taken exploiting the SPHERE apodized 
Lyot coronagraph (ALC). The diameter of the coronagraphic mask used is 185 mas. A bar indicating 
the length of 0.1 arcsec has been superimposed to the image to make the scale clearer.\label{waffle}}
\end{center}
\end{figure}

The aim of the SD algorithm is to reduce the speckle noise. This method is described 
in detail in \citet{Sparks02} and \citet{Th07}. It can be summarized by the following
steps:
\begin{itemize}
\item each single frame in the datacube is rescaled according to its wavelength in such a 
way that the speckle pattern for the star becomes similar at all wavelenghts while its companion 
is located at different positions for different frames;
\item a polynomial fit is performed along the wavelength direction for each pixel of the
rescaled datacube;
\item the obtained fit is then subtracted from the rescaled datacube to reduce the speckle noise; and
each frame of the datacube is rescaled back to the original dimensions in such a way
that the companion image is set back to the same position in all the frames.  
\end{itemize}
The main difficulties of the SD method for IFS is to properly estimate the center of each
monochromatic image in the datacube and to rescale the images because the rescaling factor 
slightly varies with wavelength due to imperfect wavelength calibration.
To overcome these problems, satellite spots were artificially introduced by  
generating periodic phase offsets on the deformable mirror as proposed in \citet{Lan13} for what concerns their use
for SPHERE observations. The use of satellite spots was previously proposed by \citet{Siva06} and by \citet{Marois2006}. 
These satellite images, symmetric with respect to the central 
star as shown in Figure~\ref{waffle}, are used to compute the star center 
position very accurately (with a precision of some tenths of spatial pixel) and, moreover, to calibrate  
the rescaling factor more accurately by exploiting their shifting outward with increasing wavelengths. 

A second order of problems consists in the fact that the image could not be uniformely 
illuminated because of an out of focus ghost on the telescope simulator optics. To solve this problem, we 
performed a bidimensional fit on the original image. 
The image resulting from this procedure is an inclined plane function that is subtracted
to the original image. While this problem arises from the telescope simulator, we cannot exclude that a similar problem could
arise for the on-sky data. For this reason this procedure should be maintained in the pipeline and used just in case it is needed.
Besides, when rescaled properly according to their wavelength, the speckle FWHM varies with 
wavelength because it is not uniquely determined by diffraction. This results in an
inaccurate subtraction of the speckles. To solve this problem, all the images
are smoothed so that the speckle FWHM remains unchanged for all wavelengths. This 
method enables more precise speckle subtraction, but the smoothing slightly degrades the spatial resolution to 
about 1.5 $\lambda /D$. Finally, to compensate the flux variations within original images and not 
properly corrected by the subtraction of the inclined plane function 
described above, each image is divided in eight radial sectors for which 
normalization factors are calculated using the median of the image for separation ranging from 6 
to 15 $\lambda /D$. The images were then divided by these normalization factors taking care to
maintain comparable flux for all the images at different wavelengths before applying the SD procedure. 
These procedures generate images more uniform in flux 
and better aligned with the speckle pattern. As a final step, we developed the SD procedure 
to exclude the planet signal when fitting the function to 
be subtracted from spectra to reduce the speckle noise. This was done by excluding all the pixels 
around the maximum of the spectrum from the spectrum to be fitted if larger than its 
median plus fifteen times the standard deviation of the remaining part of the spectrum itself. 
The number of excluded pixels is calculated according to 
the fraction of the spectrum that would be covered by a possible planet at the separation 
under consideration according to the theory of the SD given in~\citet{Th07}. 

\subsection{The PCA for IFS}\label{PCAIFS}
As an alternative to the SD algorithm described above, we implemented a 
PCA procedure to be applied to the IFS data as described, e.g., 
in~\citet{So12}. In this case, the data prereduction is identical to what has been described in 
Section~\ref{specdeconvIFS}, but the normalization of the images is performed using the image median
between 6 and 15 $\lambda /D$, without division in sectors. A 
stellar profile, computed from the median of all the pixels equidistant from the 
central star, is then subtracted from each image. Only the central part of these resulting images is 
considered to fill the two-dimensional array needed to feed the PCA procedure. The 
size of these subimages can be selected by the user, but from our experience we have found that the optimal size is
200 pixels centered on the center of the star. This allows us
to work on the most important part of the image near the center. Moreover, discarding the most external part 
of the image, we are able to avoid a procedure that is too expensive computationally. The PCA, relying on the singular value
decomposition algorithm, generates three arrays which, combined in the 
correct way, lead to the eigenvectors and the eigenvalues used to reconstruct the original 
data. The principal components for this reconstruction are given by the matrix product of 
these two arrays. A principal components subset is used to generate 
an image with the quasi static noise pattern that can then be subtracted from the original image. 
The number of principal components used is adjustable by the user. These adjustments will
be discussed in more detail in Section~\ref{resultpca}. Clearly, the 
larger the number of principal components used, the better the noise subtraction, but, 
on the other hand, the larger the cancellation of the signal from possible faint companion 
objects. 

\subsection{Angular differential imaging}\label{adi}
During the laboratory tests continuous FOV rotation by the derotator was not possible because this would result in a 
misalignment of the Lyot stop when using the TSIM since the telescope pupil is inside the TSIM itself. 
So, strictly speaking, no implementation of 
the ADI technique was possible. We tried to evaluate the efficiency of this algorithm, however, by subtracting images taken at 
different epochs. To this aim, we implemented a very simple ADI procedure that subtracts from 
each image all the other images in the datacube excluding just the adjacent images. Each
image was normalized before the subtraction dividing for the value of the total flux of the image itself. 
The analysis was performed using the data reduced with the speckle subtraction methods described 
in Section~\ref{specdeconvIFS} and~\ref{PCAIFS}. In this configuration, SPHERE was proven to be 
extremely stable, with no significant decorrelation of the instrumental speckle pattern over 
several tens of minutes. Given the great stability of the speckle pattern in the laboratory images, which will not be reproduced 
for the on-sky data, these results give an optimistic estimation of the on-sky 
instrument performances, where we expect more rapid decorrelation of instrumental speckles with timescale of about ten 
minutes \citep[see][]{Mar13} with derotation. It is important to stress that the temporal stability
will be very important for on-sky operation. Given that SPHERE was carefully designed with this aim in mind, 
however, we expect very good stability even for on-sky data. 

\subsection{Algorithm throughput}\label{cancellation}
To properly evaluate the contrast level achievable by exploiting the speckle subtraction
method described above we have to account for the algorithm throughput.
This effect is inherent to the differential techniques used in high-contrast imaging. 
In fact, these techniques require the subtraction of two images: in one image
the faint companion signal is present, and in the other it is either absent or weak. 
The two images should have a noise pattern, mainly due to speckles, as similar as possible to each other. This
leads to select images with small differences in time for angular differential
imaging or in wavelength for spectral differential imaging. The closer the images are in
time and wavelength, the more likely the faint companion signal is present
in the comparison image. Hence, once the two images are subtracted from each other, part of the
signal from the faint companion is canceled together with noise. To optimize the detection, the S/N of the companion signal should be optimized. 
This optmization is typically achieved neither when the companion signal is not affected by any
self-subtraction, nor when the speckle noise is best subtracted, leading to strong self-subtraction,
but for some intermediate self-subtraction. \par

To evaluate the impact of the algorithm throughput, we injected in the scientific datacubes simulated planets at different separations 
(0.2, 0.3, 0.4, 0.5, 0.6, 0.7 and 0.8 arcsec). To be able to clearly determine the optimal algorithm throughput for the 
single planets a wide J-band contrast of 3$\times$$10^{-4}$ was used. The contrast value was evaluated using 
the off-axis PSF not affected by the coronagraph.
The spectrum of the simulated planets was assumed to be gray. In this case, we assume that the simulated companions do not have 
identical spectra to that of the central star. It is important to point out that the algorithm used to subtract the speckles
makes use of spectral information, and the throughput is dependent on the spectral properties of the
simulated planets~\citep[see e.g.][ for the case of dual band image]{Maire2014}. Moreover, the planet flux, which is quite bright in 
our case, could impact the spectral fitting. \par
To simulate the rotation of the FOV, we inserted the simulated planets with rotation of $5^{\circ}$ between two successive datacubes 
leading to a full FOV simulated rotation of $75^{\circ}$ for 16 datacubes. The rotation considered here is of the order of $1^{\circ}$ per 
minute corresponding at Paranal to the field rotation for a target with a declination between $-10^{\circ}$ and $-38^{\circ}$ and observed during 
its transit at the meridian. A continous rotation of $5^{\circ}$ would of course introduce a smearing of a point source especially at the edge 
of the FOV. However, for this exercise we just wanted to check the speckle subtraction obtained from images taken at a distance of 5 minutes and
a rotation of $5^{\circ}$. We, then, did not consider individual single exposures, rather we took a stack of the images at different times. 
On the other hand, in the real case the ADI algorithm should be modified to take  the large smearing effect introduced
by the rotation of the FOV into account.


\section{Results for IFS YJ-mode}\label{result}

In this section we present the IFS data reduction for its YJ-mode, using the procedures detailed in
Section~\ref{datared}. An overview of the data used for this analysis is given in 
Table~\ref{table:data}.

\begin{table}
\begin{minipage}{\columnwidth}
\caption{Summary of data used.}             
\label{table:data}
\renewcommand{\footnoterule}{}       
\centering                          
\begin{tabular}{c c c c}        
\hline\hline                 
                                                                     & YJ         &  YH \\    
\hline                        
 Date                                                                & 02/05/2013 & 10/24/2013 \\      
 DIT\footnote{Detector Integration Time}                             &  4 s       & 2 s        \\
 NDIT\footnote{Number of frames for each datacube}                   &  56        & 50         \\
 Neutral Density Filter                                              &  0.0       & 1.0        \\
 N datacubes                                                         &  16        & 10         \\
\hline                                   
\end{tabular}
\end{minipage}
\end{table}

\subsection{Results obtained with the updated spectral deconvolution procedure}\label{upspecdeconv} 

\begin{figure}
\begin{center}
\includegraphics[width=8.0cm]{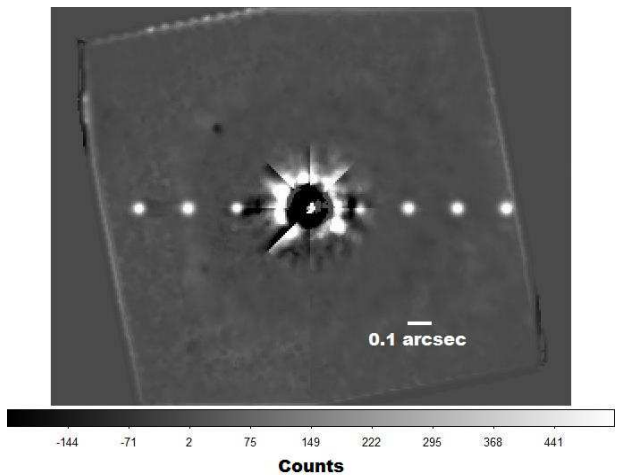}
\caption{Final reduced image obtained from the data set using the updated SD procedure where seven fake planets with contrast of 3$\times$$10^{-4}$ 
were inserted. \label{specdeconvimg}}
\end{center}
\end{figure}

We obtained very good speckle noise subtraction using the updated version of the SD and evaluating the attenuation of the satellite spots
described above. We obtained 
attenuation factors ranging from 30 to 50, depending on the function used to 
reproduce the speckle spectra. This factor is similar to the results obtained 
by~\citet{Th07} and~\citet{Crepp11}. However, 
to estimate the gain in limiting contrast, the algorithm throughput on the companion's after the SD procedure must be 
considered as described in Section\ref{cancellation}. The final reduced image obtained from the considered data set where simulated planets were inserted is displayed in 
figure~\ref{specdeconvimg}. In this Figure, the segmentation near the star center is
generated by the normalization performed in different sectors as described in 
Section~\ref{specdeconvIFS}. In the external part of the image, the data reduction method is very efficient at removing these artifacts
unlike at close separation where the method is less effective. 

\begin{figure}
\begin{center}
\includegraphics[width=8.0cm]{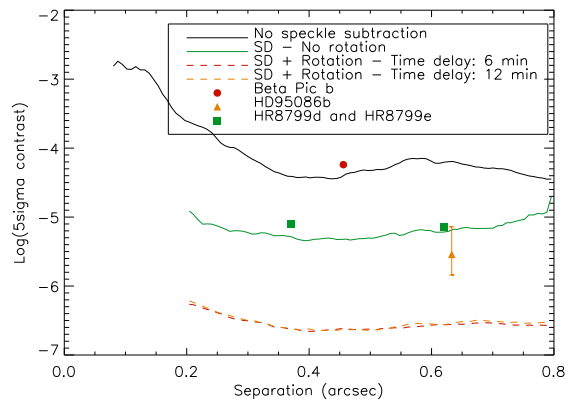}
\caption{5-$\sigma$ contrast plot with the IFS operating in YJ mode while simulating 
observations of a bright star and using the updated SD method. The contrast solid curve in black is obtained without any speckle subtraction while 
the green curve is obtained applying the updated SD procedure. The red and orange dashed curves show the contrast obtained with the simulated ADI 
procedure in addition to the speckle subtraction. 
Different symbols are used to indicate the positions of known planets on this plot. \label{YJnond_newsd}}
\end{center}
\end{figure}

In Figure~\ref{YJnond_newsd} we show the $5\sigma$ contrast curves obtained by applying
the refined SD procedure described in Section~\ref{specdeconvIFS} and 
taking the algorithm throughput into account. In the following, the contrast curves are calculated using the median image from the
final datacube produced by the speckle subtraction procedure. 
The exact procedure used to calculate the contrast follows these steps~\citep[see e.g.][]{Mas2005}:
\begin{itemize}
\item The standard deviation inside a 1.5$\lambda /D$ wide box centered on each pixel is calculated for each pixel of the image.
\item A median of these standard deviations is calculated for all the pixels located at the same separation from the central star.
\item The contrast at different separations is then calculated by dividing these median values by the
normalization factor. The normalization factor is obtained by multiplying the maximum of the off-axis PSF, derived from the star
not affected by the coronagraph, the ratio between the coronagraphic image single exposure and the off-axis image exposure, and 
the ratio of the transmission of the neutral density filters used in the two cases to avoid the saturation of the detector.
\end{itemize}

The same method was used in~\citet{Esp13} for the LBT observations of HR8799 system giving good results when comparing the S/N in the final image with 
the S/N obtained from the contrast plot. As an example HR8799e was just above the detection limit in our contrast plot confirming the fact that it 
was clearly visible in the final image. Although it provides a good view of the contrast performance, this is not the most used method 
to calculate the contrast plot, which is usually obtained by deriving the standard deviation along a ring around the center of the image for each
distance to the center. For this reason, we have dedicated further discussion to comparing the two methods in Appendix~\ref{app-contrast}.  

Figure~\ref{YJnond_newsd} shows the final results of this procedure. The contrast solid curve in black is obtained from the original coronagraphic images, the green curve is obtained from the 
application of the SD and the orange and red dashed curves represent the contrast after the application of the simulated ADI procedure. These contrasts were obtained by applying the SD and 
fitting a 1 degree function to obtain a good speckle subtraction while keeping the algorithm throughput to an high level. In this figure we do not show the contrast curves corresponding to a separation lower than 0.2 arcsec (smaller than the 
{\em bifurcation radius} as defined by \citealt{Th07}) from the central star because at these separations the method is no more effective. On this plot we added the 
positions of some known planets to enlighten the capability of IFS to retrieve their signal. Since the J magnitude for HD95086b is not
known we have exploited its known H-K color-color diagrams to infer their possible magnitudes assuming the same J-H index. The error bar displayed in the figure account for the
uncertainties of the method. We can see that we will be able to image $\beta$Pic b just using the SD and without applying the ADI
while the planets of HR8799 are close to the limit for speckle deconvolution alone. 
On the other hand, we will have to use the ADI to be able to image HD95086b. 

\begin{figure}
\begin{center}
\includegraphics[width=8.0cm]{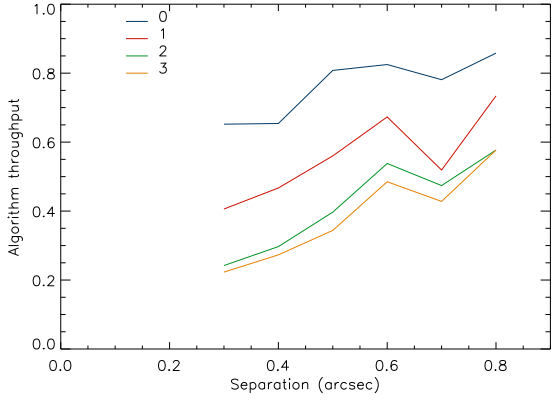}
\caption{Algorithm throughput obtained for various degrees of the fitting function used in the 
SD procedure. A detailed explanation of the meaning of these curves is given in the text. \label{figcancellation}}
\end{center}
\end{figure}

The algorithm throughput obtained for each planet is given in Figure~\ref{figcancellation} for fitting function (used in the SD procedure) degrees of 0 (corresponding to a constant value 
given by the median of the spectra), 1, 2 or 3. In this figure and in the following, 
the algorithm throughput has a value of 1 for no flux loss of the fake planet while it has a value of 0 when the planet is completely
canceled out. Clearly the throughput is much larger with lower fitting degrees. Besides, the difference between higher degrees (2 and 3) is
smaller than the difference between lower degrees. Furthermore, as expected, the 
throughput is larger at larger separations. This occurs because companion objects extend 
on smaller fractions of the speckles field spectrum at large separations (see~\citealt{Sparks02}).

\begin{figure}
\begin{center}
\includegraphics[width=8.0cm]{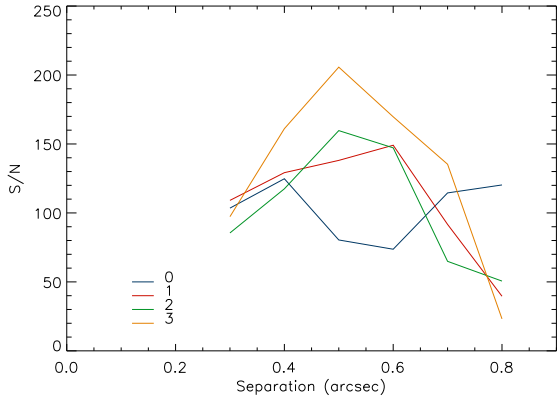}
\caption{S/N plots obtained with different degrees of the fitting function used in the SD procedure. A detailed 
explanation of the meaning of the plotted lines is given in the text. \label{sn}}
\end{center}
\end{figure}

However, high-degree fitting functions provide a better reduction of the noise.
For detection purposes, the best way to perform the SD is to optimize the S/N. 
These plots are shown in Figure~\ref{sn}. The S/N is obtained by dividing the planet signal integrated within a 1 pixel radius aperture
centered on the planet position, 
by the standard deviation estimated in a 10$\times$10 pixels region located near the considered planet but at a separation such 
that the flux in this region is not affected by the planetary signal. This is an acceptable way to estimate the noise for our images 
since that the noise is very stable in the whole image. Figure~\ref{sn} shows that at separation lower than 0.35 arcsec 
and larger than 0.75 arcsec, a fit with a low-degree polynomial function works better, while for intermediate separations the fit with a 3 degrees fitting function gives the best S/N. 
Edge effects are caused by the noisy structure at the edge of the IFS FOV that can be seen, e.g., in Figure~\ref{waffle} or 
in Figure~\ref{specdeconvimg} and are important at separations greater than 0.6 arcsec. They can explain the steep fall of the S/N value 
at large separations obtained with high-order fits.

\begin{figure}
\begin{center}
\includegraphics[width=8.0cm]{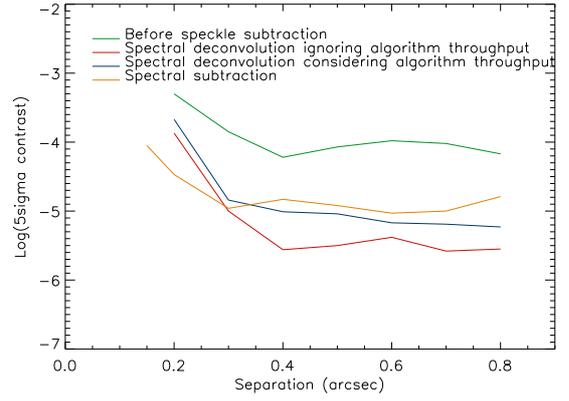}
\caption{Contrast plots for the SD taking and not taking into account the algorithm throughput.\label{contrastsd}}
\end{center}
\end{figure}

Figure~\ref{contrastsd} displays the $5\sigma$ contrast curves obtained by taking 
the SD algorithm throughput (blue line) into account compared to the contrast obtained without taking it 
(red line) into account and before the application of the SD (green line). 
The difference from Figure~\ref{YJnond_newsd} is that these contrasts are obtained
just considering the simulated planets. Despite this difference, the two latter curves are in good agreement with those
shown in Figure~\ref{YJnond_newsd} at least at separations larger than 0.3 arcsec. The discrepancy at 0.3
arcsec can be given by the proximity to the central star and to the lower throughput of the SD at that separation.
We have then tried to implement a very simple simultaneous spectral differential imaging (S-SDI) procedure by subtracting 
images obtained at different wavelengths after simulating a T7 spectral type planet. 
We have used the images at wavelengths between 1.10 and 1.19 micron where the signal is not present
to be subtracted from the images at wavelengths in the ranges between 1.02 and 1.09 micron and between 1.22 and 
1.29 micron where the signal from the planet is instead present.
In this figure the contrast curve obtained by applying this method
is represented by the orange curve. The S-SDI is more effective at small separations from the star 
(less than 0.3 arcsec) where the SD cannot be properly applied because the 
planet spectrum covers the entire spectrum provided by the IFS (see the definition of {\em bifurcation 
point} in \citealt{Th07}). 

\subsubsection{Synthetic planets spectral retrieval}\label{specrec}

\begin{figure}
\begin{center}
\includegraphics[width=8.0cm]{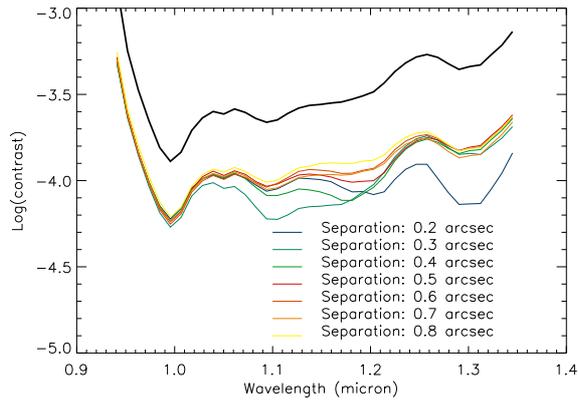}
\caption{Output spectra for the 7 simulated planets obtained applying just the spectral 
deconvolution procedure. The black solid line represents the contrast for each IFS
channel obtained from the input spectrum of the simulated objects.\label{spectrasd}}
\end{center}
\end{figure}

\begin{figure}
\begin{center}
\includegraphics[width=8.0cm]{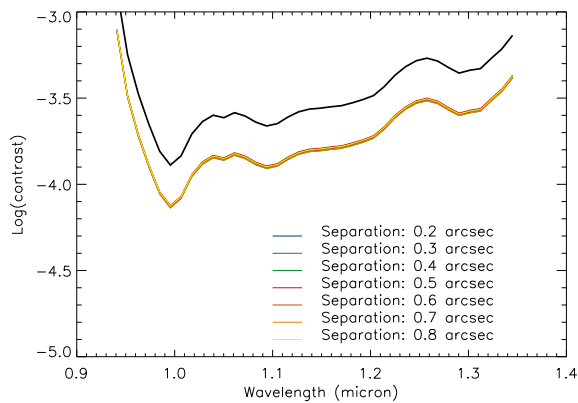}
\caption{Output spectra for the 7 simulated planets obtained applying the spectral 
deconvolution procedure and the simulated ADI procedure. The black solid line represents
the contrast for each IFS channel obtained from the input spectrum of the simulated 
objects.\label{spectrasdadi}}
\end{center}
\end{figure}

We performed a test to check the ability of the system to retrieve the spectra of the 
injected planets. The test was performed using the same data described in the previous 
paragraphs. The data were reduced using both the updated SD procedure alone
and together with the simulated ADI procedure described in Section~\ref{adi}. The injected 
simulated planets had a mean contrast on the whole wavelength range of the IFS YJ-mode of 3$\times$$10^{-4}$.
While the flux of the simulated planets was the same for all the instrument channels, the flux for the simulated star was
variable and for this reason the contrast of the simulated planets changed for different channels. 
The output spectra for the planets at different separations are shown in Figure~\ref{spectrasd} for the
case in which we only used the SD procedure. In 
Figure~\ref{spectrasdadi}, however, we only used the ADI procedure. The latter gives 
a better reconstruction of the spectra and a lower dispersion of the results. 
In this second case all the spectra are very similar to each other indipendent from the
separation from the central star. This is probably due to the high contrast of the simulated planets
that we have used in this case.
However, in both cases
we found a loss of flux with respect to the original spectra (represented by the black
solid line in both the Figures). The final flux is around 44\% of the original in the case of
the spectral deconvolution alone, while it is of the order of 57\% when we use the ADI procedure.
This is due to the low algorithm throughput given by the speckle 
subtraction methods we describe.\par
A more refined analysis of the spectro-photometric capabilities of the IFS and IRDIS and a description
of the methods adopted to improve the algorithm throughput is reported in \citet{Zurlo14}. In that work the effects of different types of
input spectra for the simulated planets are enlightened too.  

\subsection{Results with the PCA procedure}\label{resultpca}

\begin{figure}
\begin{center}
\includegraphics[width=8.0cm]{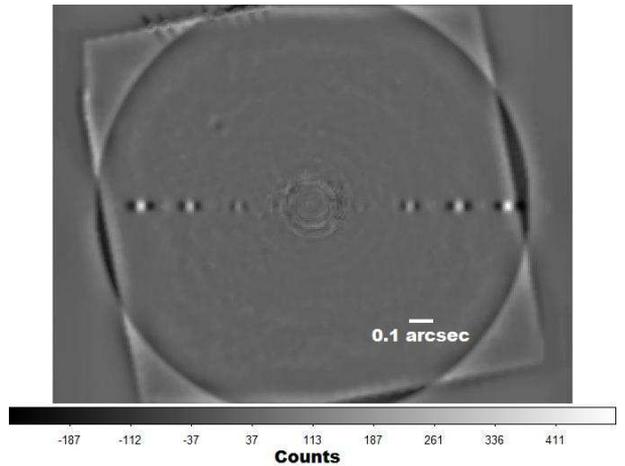}
\caption{Final image obtained by applying the PCA method to a dataset containing
simulated planets with a contrast of 3$\times$$10^{-4}$. The image is obtained
retaining only the first 4 principal components out of 38 and without considering the 
rotation of the FOV.\label{pcaimage4}}
\end{center}
\end{figure}

\begin{figure}
\begin{center}
\includegraphics[width=8.0cm]{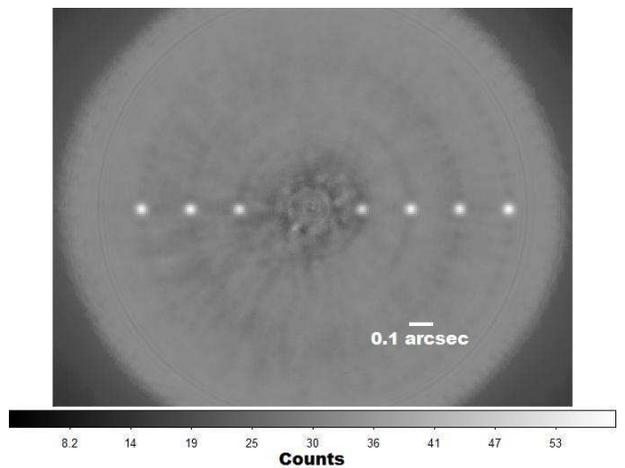}
\caption{Same as for Figure~\ref{pcaimage4} but in this case the image is obtained
retaining the first 32 principal components out of 608 and considering the rotation of the FOV 
too. Moreover, in this case the simulated planets have a contrast of $10^{-5}$.\label{pcaimage32}}
\end{center}
\end{figure}

\begin{figure}
\begin{center}
\includegraphics[width=8.0cm]{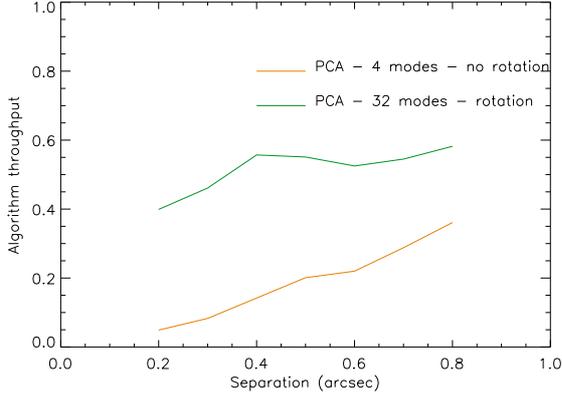}
\caption{Algorithm throughput versus separation for two sets of the PCA method parameters. The
algorithm throughput is represented by the orange curve and corresponds to no rotation of the FOV and
4 principal components, as shown in Figure~\ref{pcaimage4} while the 
green line represents the FOV rotation case and 32 principal components 
as shown in Figure~\ref{pcaimage32}.\label{cancpca}}
\end{center}
\end{figure}

Figure~\ref{pcaimage4} shows the final image obtained by applying the PCA procedure on a 
single science datacube without considering rotation of the FOV. Similar to the SD case, 
seven fake planets with 3$\times$$10^{-4}$ contrast were injected at the same separation as above.
In this figure negative features along the direction star-planets arising from the subtraction procedure are 
clearly visible. They are caused by the oversubtraction given by a not perfect spectral fitting around the planetary position.
Figure~\ref{pcaimage32} displays the final image obtained using all the datacubes and where the rotation of the FOV is 
simulated with a rotation of $5^{\circ}$ between consecutive datacubes corresponding to a total FOV rotation
of $75^{\circ}$. Here we have decided to insert simulated planets with a lower contrast of $10^{-5}$, which
is nearer to the contrast limit for this image and has a smaller impact on a correct determination of the S/N probably because
of the lower level of the strong subtraction structures described below. 
In the latter case, the image has been reconstructed using 32 principal components 
(see Section~\ref{PCAIFS}), while in the former case, without FOV rotation, only four principal components were used.
Negative structure given by the subtraction algorithm are visible in Figure~\ref{pcaimage32}. 
The strong level of the noise suppression that we can obtain with this technique 
combined with the very high contrast of the simulated planets enables these structure to be visible. 
In future improvements of this procedure, we want to address in detail these residual structures in the image.
However, the presence of these structures does not affect the scientific results presented in this paper.   
Similar to the SD case, we exploited these images to calculate the algorithm throughput 
at different separations from the central star. The algorithm throughput versus the 
separation from the central star is shown in Figure~\ref{cancpca}.

\begin{figure}
\begin{center}
\includegraphics[width=8.0cm]{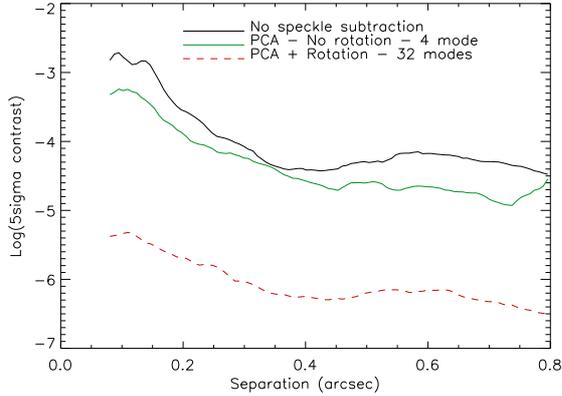}
\caption{5$\sigma$ contrast plot with the IFS operating in YJ mode for bright star observations after PCA data reduction and with simulated  
ADI procedure. The solid black line represents the contrast before speckle subtraction, 
the green line represents the contrast after PCA data reduction method using only the 4 principal 
components and without any rotation (see Figure~\ref{pcaimage4}).  
The red line is obtained applying the PCA method with 32 principal components and with 
simulated FOV rotation (see Figure~\ref{pcaimage32}). \label{YJnond_pca}}
\end{center}
\end{figure}

Using these algorithm throughputs, the contrast plot displayed in Figure~\ref{YJnond_pca} were properly estimated. In this
Figure the solid black line represents the contrast without any speckle subtraction and the solid green line represents the contrast 
without field rotation (4 principal components out of 38 are used). The image resulting from the complete 
reduction of this data set is shown in Figure~\ref{pcaimage4}. 
The dashed red line represents the contrast for the case where the FOV rotation is simulated and 32 
principal components out of 608 are used. The final image obtained from this data set is shown 
in Figure~\ref{pcaimage32}. 

\begin{figure}
\begin{center}
\includegraphics[width=8.0cm]{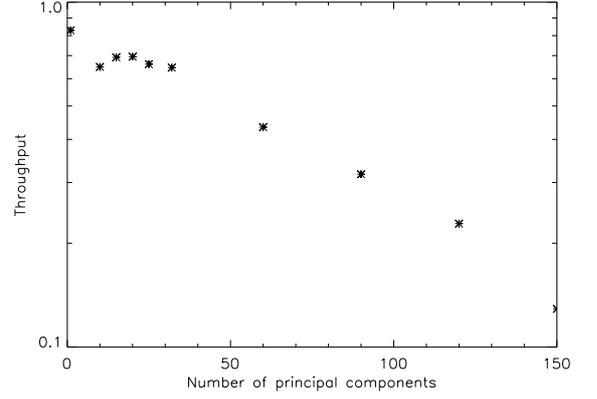}
\caption{Algorithm throughput by the PCA method for a planet at 0.4 arcsec using 
different numbers of principal components.\label{pcacanc}}
\end{center}
\end{figure}

\begin{figure}
\begin{center}
\includegraphics[width=8.0cm]{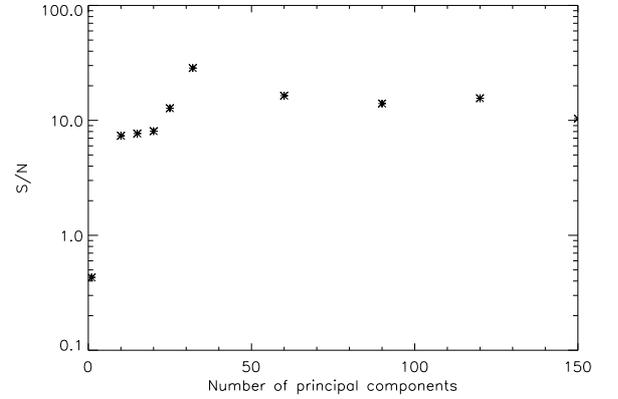}
\caption{S/N for the PCA method with a planet at 0.4 arcsec using 
different numbers of principal components.\label{pcasn}}
\end{center}
\end{figure}

We also focused on the variation of the S/N and of the algorithm throughput with the 
number of principal components used to reconstruct the original data. The throughput and S/N values for the planet at 0.4 arcsec separation are given 
in Figure~\ref{pcacanc} and in Figure~\ref{pcasn}, respectively. As expected, the throughput reduces with the number of principal components used, but at the same time
the noise decreases so that the S/N is maximum when 32 principal components are used. In order to find faint companions, 
the number of principal components to be used needs to be carefully considered 
since it changes from case to case and, for instance, it depends on the separation from the 
star~\citep[see e.g.][]{Meshkat2014}. However, for this analysis 
with simulated planets of known contrast we preferred to use a single number of principal components for the whole image. This choice
is adequate for the objective of this analysis, which is mainly aimed at providing an initial estimate of the contrast that can be reached with this 
procedure. Moreover, this choice allowed us to use a simpler code and, as a consequence, to reduce 
the required computing time.

Tests made using different numbers of principal components have allowed to us to find that it is 
adequate to use 32 of them as shown in Figure~\ref{YJnond_pca}. We would like to stress that this number 
is adequate just for the present data set. Different data sets will certainly require the use of a different number
of principal components that must be evaluated on a case-by-case basis.  

\section{Results for IFS YH-mode}\label{resultyh}
Following the scheme of Section~\ref{result}, we present the results of the analysis of the data taken
with IFS operating in the YH-mode. As seen in Table~\ref{table:data}, the data used 
for this instrument mode were taken at a different epoch with respect to the data taken for the YJ-
mode. Between the two epochs the condition of the instrument changed. In particular 
the telescope simulator lamp was changed and an additional neutral density filter was used to
avoid saturation. Moreover, the spectra obtained from the telescope simulator presented a strong absorbing band at 1.4 $\mu$m 
caused by the $OH^{-}$ ions contained in the optical fiber used. This caused extremely low flux 
at the wavelengths around 1.4 $\mu$m making it more difficult to apply the 
SD. For these reasons the contrast obtained in this mode is worse than 
those for the YJ-mode. However, this problem should not occur on sky since the stellar spectra
are flatter in this wavelength range with the resolving power of IFS, even considering the
absorption due to water vapour in the Earth atmosphere. 

\subsection{Results with the updated spectral deconvolution procedure}\label{upspecdeconvyh}
\begin{figure}
\begin{center}
\includegraphics[width=8.0cm]{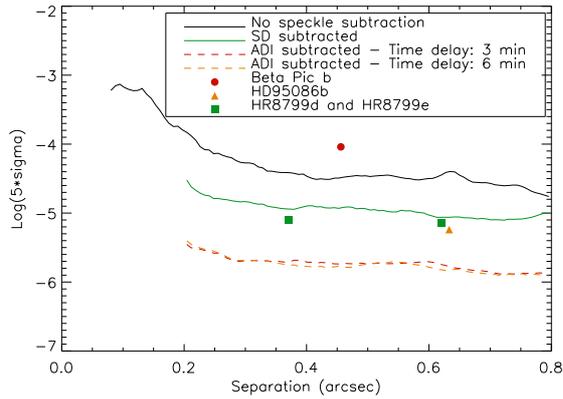}
\caption{Contrast plot obtained applying the SD and the ADI to the data taken with IFS
operating in the YH-mode. The curves color codes are the same as in Figure~\ref{YJnond_newsd}.
The difference in time delays given in the label with respect to those in 
Figure~\ref{YJnond_newsd} is due to the different way in which the data have been taken
at two epochs. See text for more detail. Different symbols are used to indicate positions 
of known planets on this plot.\label{specdeconvyh}}
\end{center}
\end{figure}

In Figure~\ref{specdeconvyh} the contrast plot obtained from the YH-mode IFS data is shown. These contrast levels are worse than those 
obtained with the instrument operating in the YJ-mode (see Figure~\ref{YJnond_newsd} for 
comparison). The worsening is particularly evident for the simulated ADI contrast 
plot while it is much less evident for the SD alone contrast plot. 
This worsening could be due, besides the reasons given above, to the shorter total integration time for the data set used in this test.    
As for YJ-mode case, we have inserted the positions of some known planets on the plot. In this case,
we will be able to image the planet of $\beta$Picb without the application of the ADI, while
to retrieve the signal from HD95086b and from the HR8799 planets, the use of ADI is mandatory.

\subsubsection{Synthetic planets spectral retrieval}\label{specrecyh}
\begin{figure}
\begin{center}
\includegraphics[width=8.0cm]{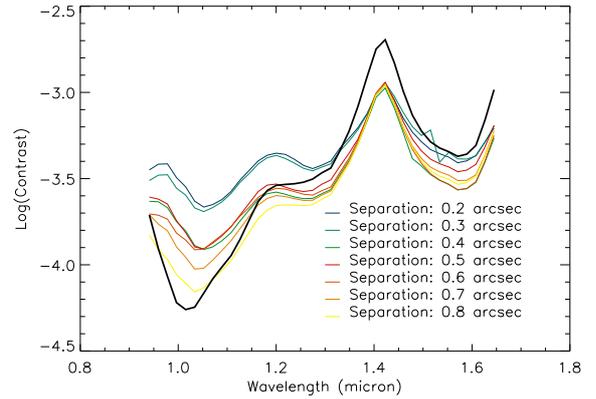}
\caption{Spectral retrieval obtained for seven planets at different separations with the IFS
operating in the YH-mode. In this case both the SD and the ADI 
were applied to the datacube before performing the spectral retrieval.
The black solid line represents the contrast for each IFS channel obtained from the input 
spectrum of the simulated objects. \label{plotspecrecyh}}
\end{center}
\end{figure}

\begin{figure}
\begin{center}
\includegraphics[width=8.0cm]{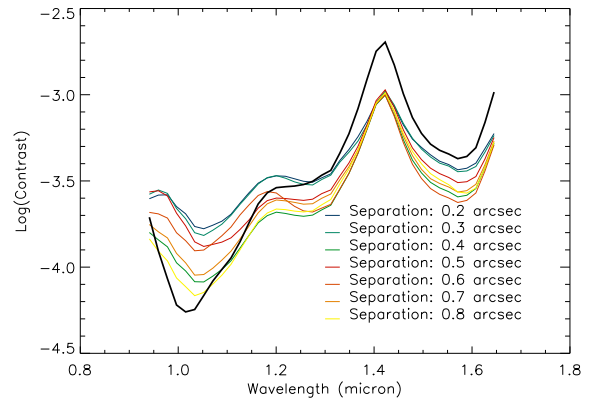}
\caption{Spectral retrieval obtained for seven planets at different separations with the IFS
operating in the YH-mode. In this case only ADI was applied to the 
datacube before performing the spectral retrieval.
The black solid line represents the contrast for each IFS
channel obtained from the input spectrum of the simulated objects.
\label{plotspecrecyhnsd}}
\end{center}
\end{figure}

Similar to the YJ-mode analysis, seven simulated planets were added to the science datacube to 
estimate the precision in reconstructing the spectra of the injected planets. 
The introduced fake planets have the same flux for all the IFS channels. As a consequence, given that
the flux of the simulated star was variable from one channel to the other, the contrast of the simulated planets
with respect to the simulated star was variable.  The mean contrast on the whole wavelength range of the IFS YH-mode is of
3$\times$$10^{-4}$. In Figure~\ref{plotspecrecyh} the results obtained with this datacube
after the application of both SD and ADI are shown, while 
Figure~\ref{plotspecrecyhnsd} shows the spectral retrieval obtained in this case 
after the application of ADI only on the original datacube. 

The spectral retrieval is better in the second case, confirming the results already obtained for the YJ-mode.
At the same time, this result confirms that the application of SD strongly distorts the original spectra. This effect is slightly decreased when the
ADI procedure is directly applied to the original datacube without applying the SD. 
Moreover, a second problem that affects the analysis for the YH mode is given by the great 
difference in flux between different images at different wavelengths in the YH datacube. 
This has the effect of inserting biases in the estimation of the normalization factors along the spectra and, in this way, 
leads to overestimate parts of the spectrum with respect to others.
From the analysis of the results, we can conclude that the distortion of the spectra is about eight 
times larger than in the case of the YJ-mode. Moreover, in some cases, additional flux is added to the original in such a way
that the flux is higher in the extracted spectra than in the simulated spectra. This could arise from the worse quality of the data in the 
YH-mode than the YJ-mode as explained in Section~\ref{upspecdeconvyh}. Hence, careful analysis of YH-mode images will be then
mandatory on on-sky data.

\subsection{Results with the PCA procedure}\label{resultpcayh} 
\begin{figure}
\begin{center}
\includegraphics[width=8.0cm]{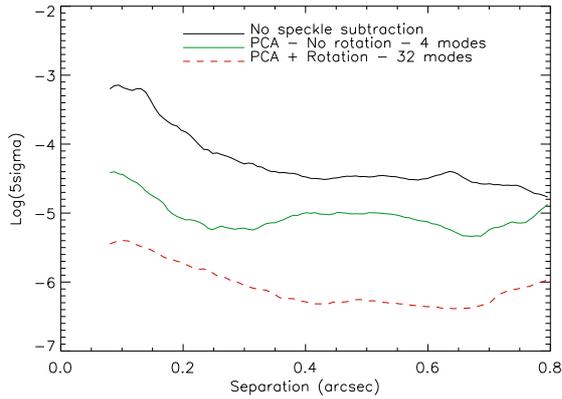}
\caption{Contrast plot obtained applying the PCA procedure to the IFS YH-data. The solid green
line represents the contrast obtained with the PCA applied to a single datacube without any 
rotation. The dashed red line represents the 
contrast obtained applying the PCA to all the datacubes with the rotation of 
the FOV. \label{plotpcayh}}
\end{center}
\end{figure}
Figure~\ref{plotpcayh} shows the contrast plots obtained applying the PCA to the IFS
YH-mode data. The green line has been obtained applying the procedure to a single datacube
without any rotation of the FOV. 
We then applied the
procedure to all ten datacubes to simulate FOV rotation as described
in Section~\ref{upspecdeconv}. The contrast plot obtained for this case is represented by 
the dashed red line in Figure~\ref{plotpcayh}. As for all previous cases, the introduction of 
the field rotation allows us to obtain a contrast level better than $10^{-6}$. 





\section{Conclusions}\label{conclusion}

\begin{table}
\caption{Summary of the contrast obtained at different separations for the IFS YJ-mode. 
The meaning of the acronyms are explained in the text.\label{table:resYJcanc}}                   
\centering                          
\begin{tabular}{c c c c c}        
\hline\hline                 
       Separation                &  SD     &  SD+ADI  &  PCA1 &  PCA2  \\    
\hline                        
 0.2                             & 1.33$\times$$10^{-5}$ & 5.67$\times$$10^{-7}$ & 1.30$\times$$10^{-4}$ & 2.51$\times$$10^{-6}$ \\      
 0.3                             & 5.94$\times$$10^{-6}$ & 3.14$\times$$10^{-7}$ & 5.74$\times$$10^{-5}$ & 1.11$\times$$10^{-6}$ \\
 0.4                             & 4.60$\times$$10^{-6}$ & 2.20$\times$$10^{-7}$ & 2.68$\times$$10^{-5}$ & 6.59$\times$$10^{-7}$ \\
 0.5                             & 4.95$\times$$10^{-6}$ & 2.36$\times$$10^{-7}$ & 2.51$\times$$10^{-5}$ & 7.52$\times$$10^{-7}$ \\
 0.6                             & 6.31$\times$$10^{-6}$ & 2.71$\times$$10^{-7}$ & 2.19$\times$$10^{-5}$ & 8.20$\times$$10^{-7}$ \\
 0.7                             & 6.88$\times$$10^{-6}$ & 2.90$\times$$10^{-7}$ & 1.45$\times$$10^{-5}$ & 5.64$\times$$10^{-7}$ \\
 0.8                             & 1.93$\times$$10^{-5}$ & 2.70$\times$$10^{-7}$ & 2.92$\times$$10^{-5}$ & 4.05$\times$$10^{-7}$ \\
\hline                                   
\end{tabular}
\end{table}

\begin{table}
\caption{Summary of the contrast obtained at different separations for the IFS YH-mode. 
The meaning of the acronyms are explained in the text.\label{table:resYHcanc}}                   
\centering                          
\begin{tabular}{c c c c c}        
\hline\hline                 
       Separation                &  SD        &  SD+ADI   &   PCA1   &   PCA2   \\    
\hline                        
 0.2                             & 3.16$\times$$10^{-4}$ & 3.32$\times$$10^{-5}$ & 8.22$\times$$10^{-6}$ & 1.90$\times$$10^{-6}$ \\      
 0.3                             & 1.02$\times$$10^{-4}$ & 1.46$\times$$10^{-5}$ & 6.04$\times$$10^{-6}$ & 7.87$\times$$10^{-7}$ \\
 0.4                             & 6.86$\times$$10^{-5}$ & 1.26$\times$$10^{-5}$ & 1.01$\times$$10^{-5}$ & 6.69$\times$$10^{-7}$ \\
 0.5                             & 6.60$\times$$10^{-5}$ & 1.13$\times$$10^{-5}$ & 9.95$\times$$10^{-6}$ & 5.96$\times$$10^{-7}$ \\
 0.6                             & 6.00$\times$$10^{-5}$ & 9.34$\times$$10^{-6}$ & 7.45$\times$$10^{-6}$ & 5.17$\times$$10^{-7}$ \\
 0.7                             & 4.73$\times$$10^{-5}$ & 8.00$\times$$10^{-6}$ & 5.47$\times$$10^{-6}$ & 5.99$\times$$10^{-7}$ \\
 0.8                             & 3.11$\times$$10^{-5}$ & 9.75$\times$$10^{-6}$ & 1.43$\times$$10^{-5}$ & 1.07$\times$$10^{-6}$ \\
\hline                                   
\end{tabular}
\end{table}

We have described the results of the functional and performance tests for the IFS instrument
within SPHERE. While the functional tests performed using the instrument templates and the DRH 
recipes allowed us to validate and improve the software, the performance tests allowed us to get 
hints about the achievable contrast with IFS when used on sky and to optimize the calibration and data reduction procedures. The spectral 
deconvolution procedure currently implemented in the dedicated DRH recipe 
yields a contrast close to $10^{-5}$, while better results were expected from instrument simulations~\citep{Mesa11}. For this 
reason we implemented an updated version of the SD procedure taking several 
issues not properly addressed in the original procedure into account(e.g., more 
precise centering of the datacube images and more refined rescaling factor at 
different wavelengths). This updated procedure allowed us to improve the contrast by a factor of 
two to three. The final contrast obtained is of the order of 4$\times$$10^{-6}$ at 
0.3 arcsec separation, taking the total throughput of the differential imaging procedures
properly into account. To consider this last effect, several simulated 
planets located at different separation from the star were injected.\par

We also developed a new procedure for the speckle subtraction exploiting the PCA algorithm.
In Table~\ref{table:resYJcanc} we summarize the $5\sigma$ contrast we
obtained with the IFS in the YJ-mode with various speckle subtraction methods described
in this paper. Similar
results are presented in Table~\ref{table:resYHcanc} for the IFS YH-mode. 
In both these Tables the second column gives the results obtained when applying the SD
to a single datacube, while in the third column lists the results obtained with our 
simulated ADI procedure after SD are displayed. In the fourth column, labeled PCA1, we list the results 
obtained applying the PCA to a single datacube and with only four principal components used 
(in this case no rotation of the FOV was simulated), while
in the fifth column, labeled PCA2, we give the results of the PCA performed on the complete data set 
in this case using 32 principal components (in this latter case a rotation of the FOV was simulated). 
While the cancellation of the noise is similar for both SD and PCA, the throughput is much lower
for the PCA method as shown in Table~\ref{table:resYJcanc}. This latter effect seems to be
much less important in the case of the YH-mode as shown in Table~\ref{table:resYHcanc}. As explained 
in Section~\ref{resultyh}, however, these data were taken under different conditions with respect 
to the YJ case and proper comparison between the two cases is difficult. Moreover this caused a 
loss of effectiveness of the SD method for the YH data (see Table~\ref{table:resYHcanc}).
All these uncertainties will be solved when on-sky data become available.\par
However, we can conclude that the PCA procedure can give results
comparable to the SD results once the algorithm throughput is properly taken into account. A more refined method
to take them into account will be presented in an associated paper \citep{Zurlo14}. Moreover,
the implementation of ADI procedure in the PCA procedure is much easier than the implementation of the ADI in the 
SD procedure. For this reason, the PCA procedure should probably be considered as a better solution for the final IFS pipeline.\par
Finally, we have verified that, whatever the choice for the IFS reduction pipeline 
(SD, PCA or other possible solutions), ADI is mandatory to reach the $10^{-6}$ 5$\sigma$
contrast threshold using IFS mounted on SPHERE. The effectiveness of the ADI, however, strongly depends on the rotation of the FOV
and on the separation from the central star.


\begin{acknowledgements}
We are grateful to the SPHERE team and all the people always available
during the tests at IPAG in Grenoble. D.M., A.Z., and S.D. acknowledge
partial support from PRIN INAF 2010 ``Planetary systems at young ages''.
SPHERE is an instrument designed and built by a consortium consisting of IPAG,
MPIA, LAM, LESIA, Laboratoire Fizeau, INAF, Observatoire de Gen\'eve, ETH,
NOVA, ONERA, and ASTRON in collaboration with ESO. 
\end{acknowledgements}

\bibliographystyle{aa}
\bibliography{ifstest_v1.1}

\begin{appendix}
\section{Contrast plot calculation}\label{app-contrast}
The method that we have used to calculate the contrast plot in this paper and described in Section~\ref{upspecdeconv} has in the past been
demonstrated to be reliable~\citep[see e.g.][]{Esp13}. On the other hand, the most common method used to calculate the contrast plot is to 
calculate the standard deviation on a ring centered on the center of the star. Our choice was driven by the fact that 
this second method does not take the possible large scale inhomogeneities in the image properly into account, and that we need however to 
calculate some local noise when we are searching for a point source like a planet. \par
\begin{figure*}
\begin{center}
\includegraphics[width=8.0cm]{plotall_flat_sb_nd0_mdn1_canc_up_cutcut_plan_big.eps}
\includegraphics[width=8.0cm]{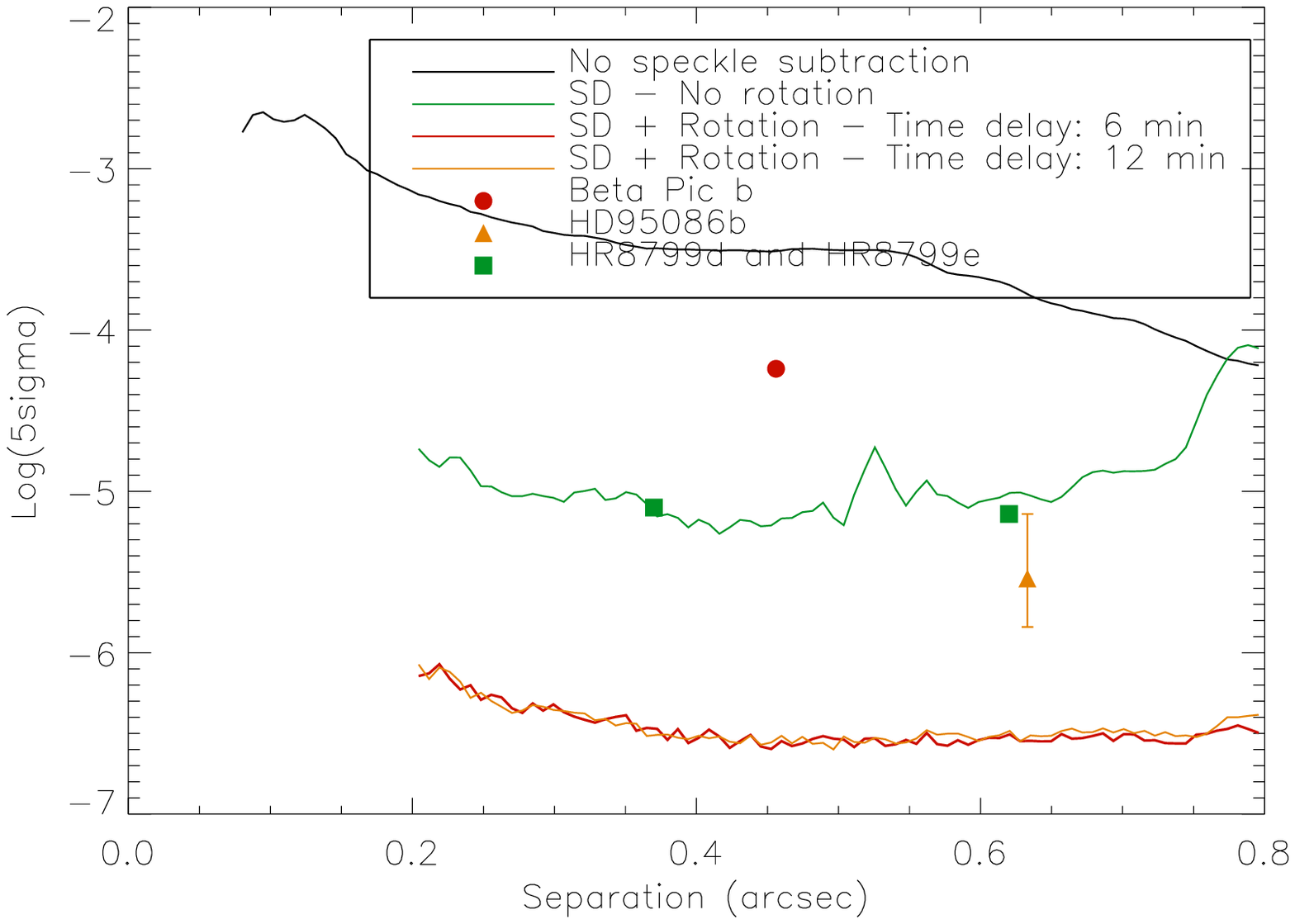}
\caption{Left: same of image~\ref{YJnond_newsd} obtained with our method. Right: same contrast plot obtained using the standard 
method described in the text. \label{comparison-nosmooth}}
\end{center}
\end{figure*}

To compare the results of the two methods we have run
both of them on one of the cases presented here. In particular, we have chosen the case presented in Figure~\ref{YJnond_newsd}. On the
left panel of Figure~\ref{comparison-nosmooth} we present the same image while on the right panel we present the contrast obtained 
using the classical method described above on the same data set. 
The difference between the two plots (black line) for nonsubtracted images is readily evident while it tends to diminish when we consider 
subtracted images and it is almost null when we consider images where the ADI is applied. The great difference for nonsubtracted images is
given by the problem of the large scale inhomogeneities in the original image cited above. To demonstrate this, we have prepared the same plots 
presented in Figure~\ref{comparison-nosmooth} but applying a low-frequency filtering on the nonsubtracted image, subtracting from 
the original image a smoothed version of the image itself using the SMOOTH IDL procedure with a large value of the smoothing area (with a
semi amplitude of 25 pixels corresponding to $\sim$0.18 arcsec). \par
\begin{figure*}
\begin{center}
\includegraphics[width=8.0cm]{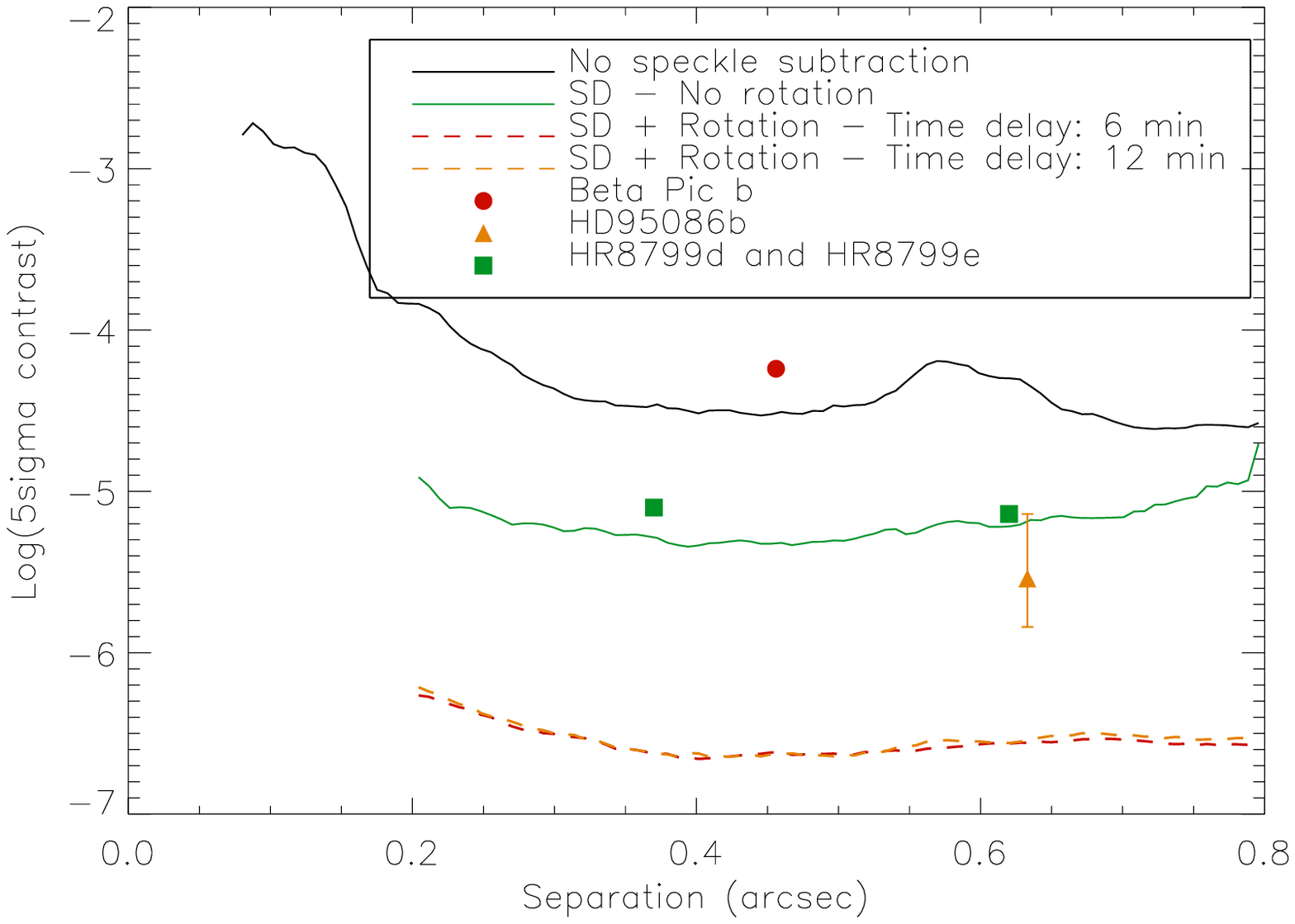}
\includegraphics[width=8.0cm]{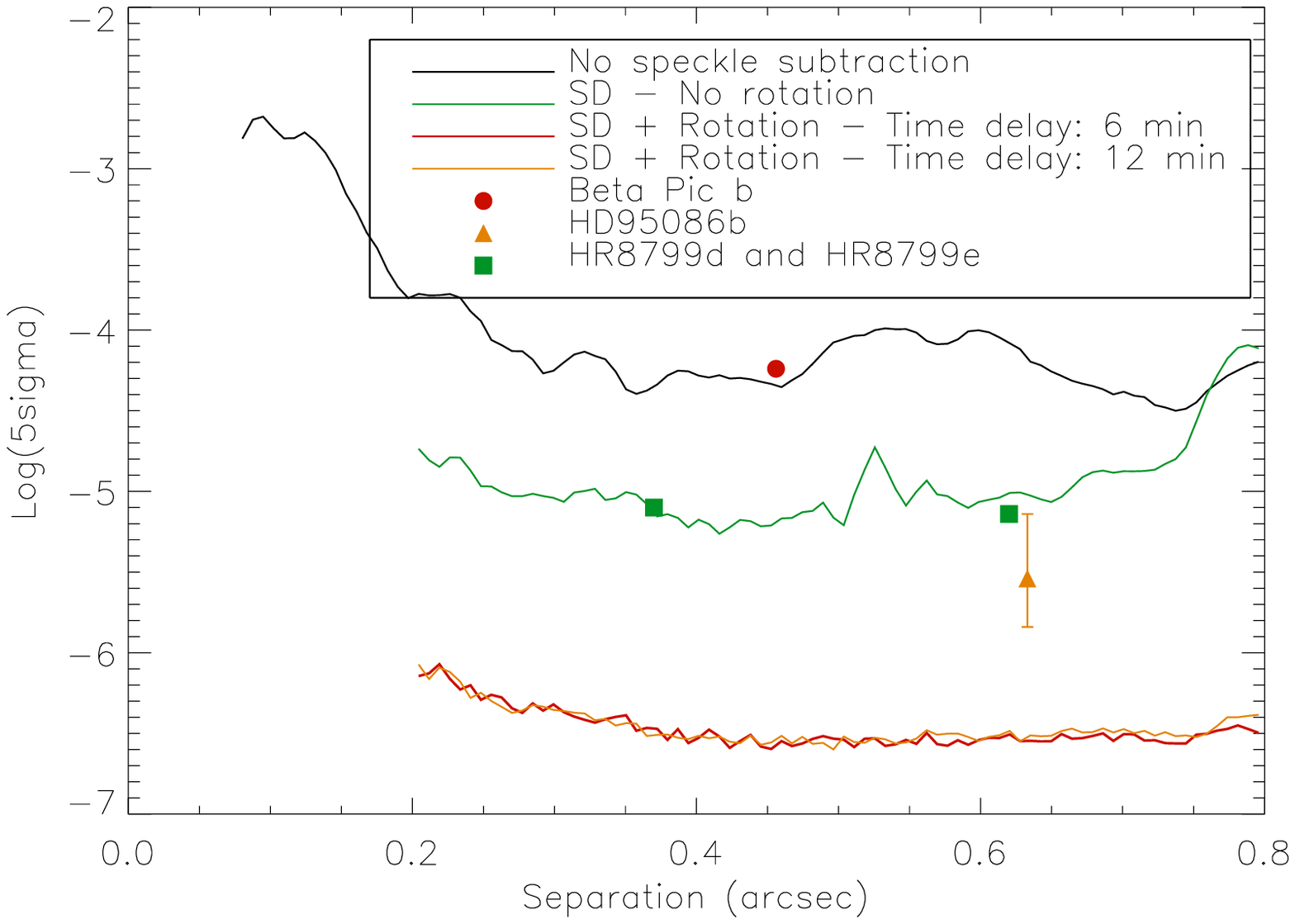}
\caption{Left: same of image~\ref{YJnond_newsd} obtained with our method where a smoothed version of the same 
image has been subtracted from the nonsubtracted image after applying a low-frequency filtering to subtract 
any possible large scale structure in the image. 
Right: same contrast plot obtained using the standard method described in the text. \label{comparison-smooth}}
\end{center}
\end{figure*}

The results of this test are presented in Figure~\ref{comparison-smooth} where it is evident that now the contrast plots are very similar 
using both methods. The conclusion of this test is that the main difference between the two methods is on the nonsubtracted images 
where the large scale inhomogeneities in the image are not taken properly into account by the standard method. These differences
can be reduced applying a low-frequency filtering on the images. The main problem with our method is that if the box that we are using for 
calculations has a dimension similar to the dimension of the speckle ($\sim$ $\lambda$/D), we could have an underestimation of the noise. However, 
we took care that the dimension of the box was larger than this value (as said in the paper, of the order of $\sim$1.5$\lambda$/D). Moreover, 
various tests with different dimensions of the box have been done arriving at the conclusion that boxes with larger dimensions do not  
significantly change the estimated contrast. From these tests, we can estimate an error of the order of 10\% for our contrast estimation. \par
\begin{figure}
\begin{center}
\includegraphics[width=8.0cm]{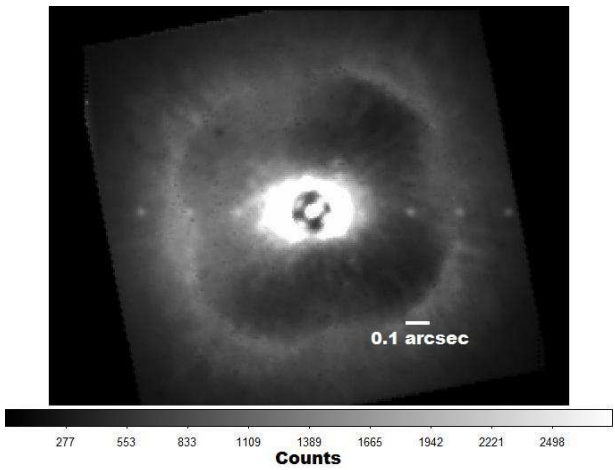}
\caption{Nonsubtracted datacube median image with simulated planets with a contrast of $10^{-4}$. The planets with separation larger than
0.2 arcsec are just visible as foreseen by our contrast plot. \label{testcontrasto}}
\end{center}
\end{figure}

 
As a further test of the reliability of our contrast estimation we have introduced simulated planets with a contrast of $10^{-4}$ 
on the nonsubtracted image (black plot on both panels of Figure~\ref{comparison-nosmooth}). The result of this test is presented in 
Figure~\ref{testcontrasto} where, as foreseen by the contrast plot, the planets at separations of 0.3 arcsec and larger are visible 
(even just above the detection limit) while the planet at 0.2 arcsec is not visible. This is in good agreement with what we can infer
from the analysis of the contrast plot displayed on the left panel and obtained exploiting our method. \par
The conclusion of these tests are that both methods can give a reliable estimation of the contrast when used in the 
correct way.

\end{appendix}

\end{document}